**Wafer-Scale Route to Quantum-Sensitive Graphene: Making Every CVD Graphene Surface an Ultrasensitive Quantum Sensor Without Cleaning or Nanostructuring**

*Yaren Sezen[1,2], Mehmet Ali Sarsil[1], Mina Sofuani[1], Rabia Pala[1,2], Onur Ergen[1,2,3,4*]*

*[1]Department of Electronics and Communication Engineering, Istanbul Technical University, Istanbul 34450, Türkiye*
*[2] Department of Nanoscience and Nanoengineering, Istanbul Technical University, Istanbul 34450, Türkiye*
*[3]Next-Ion-Energy Inc., Yuba City, CA 95991, United States*
*[5] Institute of Nanotechnology, National Research Council (CNR-NANOTEC), c/o Campus Ecotekne, Via per Monteroni, Lecce 73100, Italy*

* Dr. Onur Ergen.
**Email:** oergen@itu.edu.tr

**Author Contributions:** All authors contributed equally.
**Competing Interest Statement:** No competing interest.
**Classification:** Physical-Science

**Abstract:** Structural imperfections and processing residues in graphene are conventionally treated as defects to be eliminated before fabrication. We show they instead constitute a scalable resource: residues from incomplete removal of transfer polymers and photoresists spontaneously dewet into discrete nanodomes upon mild annealing, converting contamination into tunable quantum islands on unpatterned, as-grown CVD graphene. This requires no lithographic patterning, catalytic metals, or elevated temperatures, so any CVD graphene surface carrying typical residue can host such islands. In hourglass-shaped devices, these islands show well-defined Coulomb blockade, pseudomagnetic confinement from strain-induced hopping asymmetry, quantized charge-relaxation resistance near $h/2e^2$, and discrete capacitance steps. Coherent transport persists to room temperature, with coherence times up to 50 μs. We further demonstrate reversible, catalyst-free hydrogen detection to 20 ± 4 ppb in dry air via RF resonance shifts, matching platinum-group-metal sensitivity with no catalyst, improving on prior catalyst-free graphene sensors by three orders of magnitude. Since the mechanism depends only on residue thickness and a mild anneal step common to graphene processing, we propose this as a route to quantum-sensitive graphene: contamination need not be removed but can serve as a built-in resource turning any CVD graphene surface into a room-temperature quantum sensor without added nanofabrication.

## 1. Introduction

The established framework for solid-state quantum technologies rests on a single axiom: disorder must be eliminated. Defects, grain boundaries, and structural heterogeneities are treated as primary sources of decoherence that degrade qubit fidelity, motivating sustained investment in atomically pristine materials, millikelvin operating environments, and complex lithographic confinement. This approach has produced transformative results, superconducting transmon qubits with millisecond-scale coherence times[1,2] and spin qubits in isotopically purified silicon with gate fidelities above the fault-tolerance threshold[3,4], yet each advance deepens the dependency on conditions fundamentally incompatible with large-scale, real-world deployment. A different logic has quietly matured. Point defects in wide-bandgap semiconductors, particularly divacancy spins in 4H-SiC, have shown that structural imperfections can harbor room-temperature quantum coherence with millisecond spin dephasing times and optically detected magnetic resonance addressable by infrared light[5,6]. Near-surface alkene-terminated divacancies in 4H-SiC further demonstrate ambient-condition sensing competitive with diamond nitrogen-vacancy centers while emitting within the near-infrared biological transparency window [6], substantially lowering barriers to practical deployment. The lesson is clear: disorder is not inherently the enemy of coherence. What matters is whether disorder generates a confinement potential that isolates a quantum degree of freedom from its environment or merely scrambles it.

Graphene offers a physically distinct and arguably more scalable realization of this principle. Its linear Dirac dispersion, sub-millielectronvolt spin–orbit coupling, and negligible $^{13}C$ hyperfine interaction make it an intrinsically favorable host for long-lived charge and spin coherence.[7] Crucially, mechanical deformation of the honeycomb lattice breaks the local equivalence of the three nearest-neighbor hopping directions, generating a valley-antisymmetric vector potential, the pseudomagnetic field (PMF), that confines Dirac fermions without violating time-reversal symmetry and without any external magnet.[8] Scanning tunneling spectroscopy of graphene nanobubbles grown on platinum demonstrated pseudo-magnetic fields exceeding 300 T, sufficient to resolve fully quantized pseudo-Landau levels at room temperature.[9] Subsequent transport measurements of CVD graphene folds revealed Coulomb-blockade signatures consistent with strain-defined charge islands[10], and theoretical calculations have extended this concept to electrically tunable spin-qubit architectures in strain-induced double quantum dots.[11] Together, these results define a compelling and largely unrealized prospect: all-electrical, room-temperature quantum coherence in a CMOS-compatible material platform, achieved without cryogenic infrastructure, optical addressing, or deliberately fabricated nanostructures. The obstacle that has prevented this prospect from becoming reality has been hiding in plain sight. Scalable chemical vapor deposition (CVD) graphene, combined with standard PMMA-assisted wet transfer and I-line photoresist patterning, invariably leaves a persistent nanometer-scale organic residue layer covering every device.[12] Despite intensive efforts spanning thermal annealing, plasma cleaning, laser ablation, and AFM contact-mode sweeping, this contamination has never been fully eliminated.[13] The field has responded by treating it as irreducible noise, a performance floor to minimize rather than a material to understand.

Here we show that controlled disorder can instead serve as a functional resource, defining an original quantum device of pseudomagnetically confined islands that support room-temperature coherence. We show that organic residues arising from incomplete removal of CVD transfer polymer (PMMA) and I-line photoresist undergo spontaneous thermodynamic self-organization into discrete, spatially localized nanodomes upon mild thermal annealing (~150 °C, CMOS-compatible). Each nanodome imposes a radially symmetric lattice deformation on the underlying graphene that generates a pseudomagnetic confinement potential capable of simultaneously defining a tunnel-coupled charge-puddle barrier and a Coulomb-blockaded single-electron island, without lithographic patterning, suspended membranes, or non-standard substrates. We characterize these self-assembled quantum islands in hourglass-shaped graphene single-electron transistors using scanning gate microscopy (SGM) and broadband RF reflectometry (50–300 MHz). RF reflectometry reveals a sharp crossover from low-frequency Coulomb-blockade-dominated transport to resonant tunneling near 200 MHz, a crossover frequency set by the intrinsic charge-relaxation RC time constant of the island, where the charge-relaxation resistance Rq = $h/2e^2$ is the universal single-channel quantum[14,15] and the island capacitance C ≈ 62 fF. Quantized charge-relaxation resistance approaching $h/2e^2$, together with discrete ~1 fF single-channel capacitance steps resolved by SGM, confirm phase-coherent single-channel transport. Time-domain measurements reveal Rabi oscillations with coherence times $T_2$ spanning 6–50 μs at room temperature in a standard Si/$SiO_2$ device architecture fully compatible with CMOS back-end processes. This is consistent with our previous finding that two-dimensional heterostructures possess material-specific critical frequencies at which disordered, stochastic transport reorganizes into coherent collective dynamics.[16]

We also demonstrate the practical application of these quantum islands on hydrogen sensing. Hydrogen safety monitoring is an enabling technology for the emerging clean-energy economy, and the state-of-the-art at strict room temperature rests entirely on noble-metal catalysts: palladium, platinum, and their alloys, which introduce phase-transition instability under hydrogen loading, susceptibility to CO and sulfur poisoning, and fabrication complexity that forecloses wafer-scale integration.[17-20] The current room-temperature detection record stands at 40 ppb using a fullerene-decorated PdCo nano-resistor network[18]; the overall benchmark is 20 ppb with a Pd–Au MEMS resonator operated at 60 °C.[21] Catalyst-free graphene chemosensors have been confined to the low-ppm regime.[22] We demonstrate that the extreme charge sensitivity of a single-channel Coulomb-blockaded nanodome island enables fully reversible, catalyst-free hydrogen detection to 20 ppb in dry air at room temperature, matching the noble-metal state of the art without any catalytic component. $H_2$ physisorption on the bare graphene basal plane, which is strictly non-dissociative at room temperature (binding energy ~48 meV; chemisorption barrier > 2.6 eV [41]), modulates the local electrostatic environment of individual nanodome islands, producing measurable shifts in the RF reflectometry resonance frequency with complete baseline recovery across multiple gas cycles. This transduction mechanism, reading a quantum-limited single-channel observable rather than a bulk conductance, is fundamentally distinct from all prior graphene chemiresistor and field-effect-transistor sensor architectures.[22]

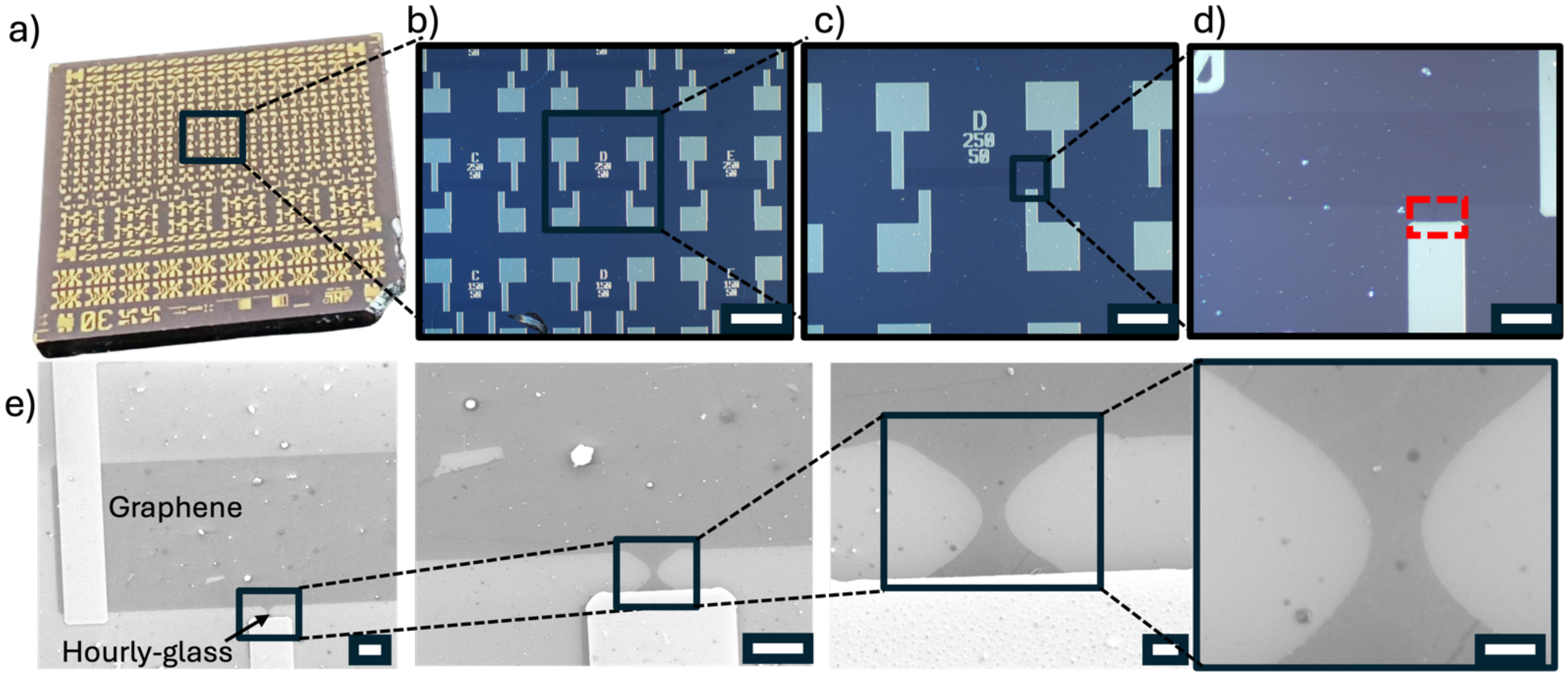

**Figure 1 Device fabrication and characterization of the hourglass graphene single-electron transistor.** (a) Optical photograph of the full chip showing the array of devices fabricated on a Si/SiO₂ substrate using standard CMOS-compatible I-line photolithography and CVD graphene transfer. (b, c) Optical micrographs at progressively higher magnification showing individual hourglass-shaped graphene devices with Au/Cr contacts, labeled by channel width (D = 250/50 nm and variants). Scale bars as indicated. (200 µm, 100 µm) (d) High-magnification optical micrograph of a single device; the red dashed rectangle indicates the hourglass constriction region containing the active graphene channel (Scale bar: 20 µm). (e) Scanning electron micrographs at progressively higher magnification showing the hourglass geometry of the graphene channel between source and drain contacts. (Scale bars: 20µm, 10 µm, 2 µm, 1 µm)

## 2. Results and Discussion

Graphene samples were obtained from Applied Nanolayers (recently acquired by Black Semiconductor) to ensure the use of high-quality, large-area CVD graphene suitable for scalable device fabrication (Fig. 1a). Multiple devices with identical geometries were fabricated on the same chip to assess reproducibility and device-to-device variability.[24,25] As shown in Fig. 1b–d, the graphene was patterned into rectangular contact regions connected via trapezoidal, hourglass-shaped channels, which serve as the primary structures examined in this study. SEM images of the hourglass channels at progressively higher magnification are shown in Fig. 1e.

Fig. 2a presents an optical micrograph of the hourglass constriction with three measurement points indicated, points (1) and (2) within the hourglass channel, and point (3) on a reference graphene strip on $SiO_2$, alongside AFM topography and phase images of the boxed region. The topography image reveals self-assembled nanodomes (bright features, confirmed height 0.54–2.7 nm by line profile analysis in Fig. 2e) distributed across the constricted channel following thermal annealing, concentrated preferentially at the channel waist. The phase image confirms distinct mechanical contrast between the graphene channel, residue nanodomes, and the underlying $SiO_2$ substrate, consistent with compositionally and mechanically distinct surface domains.

Fig. 2b presents Raman spectra acquired at points (1) and (2) within the hourglass channel (red and orange traces) and at the reference graphene strip at point (3) (black trace), all normalized to their respective maximum 2D-band intensity. Three canonical spectral features are resolved: the D band (~1344 $cm^{-1}$ at hourglass points, ~1341 $cm^{-1}$ at the reference), the G band (~1583 $cm^{-1}$), and the 2D band (~2687, 2684, and 2674 $cm^{-1}$ at points 1, 2, and 3 respectively; see Table S1). The reference graphene strip exhibits the lowest D-to-G intensity ratio ($I_D/I_G \approx 0.55$) and the highest 2D-to-G ratio ($I_{2D}/I_G \approx 2.23$), consistent with high-quality, low-disorder monolayer graphene.[26] In contrast, the hourglass channel points display markedly elevated D-band intensity ($I_D/I_G$ of 1.37 and 0.95 at points 1 and 2 respectively), indicative of enhanced intervalley scattering arising from structural perturbation concentrated at the trapezoidal constriction.[27,28] This enhancement is not attributed to bulk lattice damage but rather reflects the spatially non-uniform strain field generated by the hourglass geometry, which is known to induce pseudo-magnetic fields in graphene and promote localized intervalley scattering at the channel waist.[8,9] Despite the elevated D band, the 2D band in both hourglass spectra remains a single, well-resolved peak, confirming that the $sp^2$ carbon network is structurally preserved.[26,29] A raised spectral baseline in the 2800–3200 $cm^{-1}$ region of the hourglass trace is attributed to localized residual contamination, polymer residue or adsorbates, confined at the narrowest point of the channel by the strain-induced pseudo-magnetic potential (Table S1).[9-12] Taken together, the Raman data provide direct spectroscopic evidence that the trapezoidal geometry selectively concentrates disorder, strain, and residual contamination at the channel center, converting an otherwise disregarded perturbation into the physical basis for the electrostatically tunable graphene quantum dot characterized in subsequent measurements.

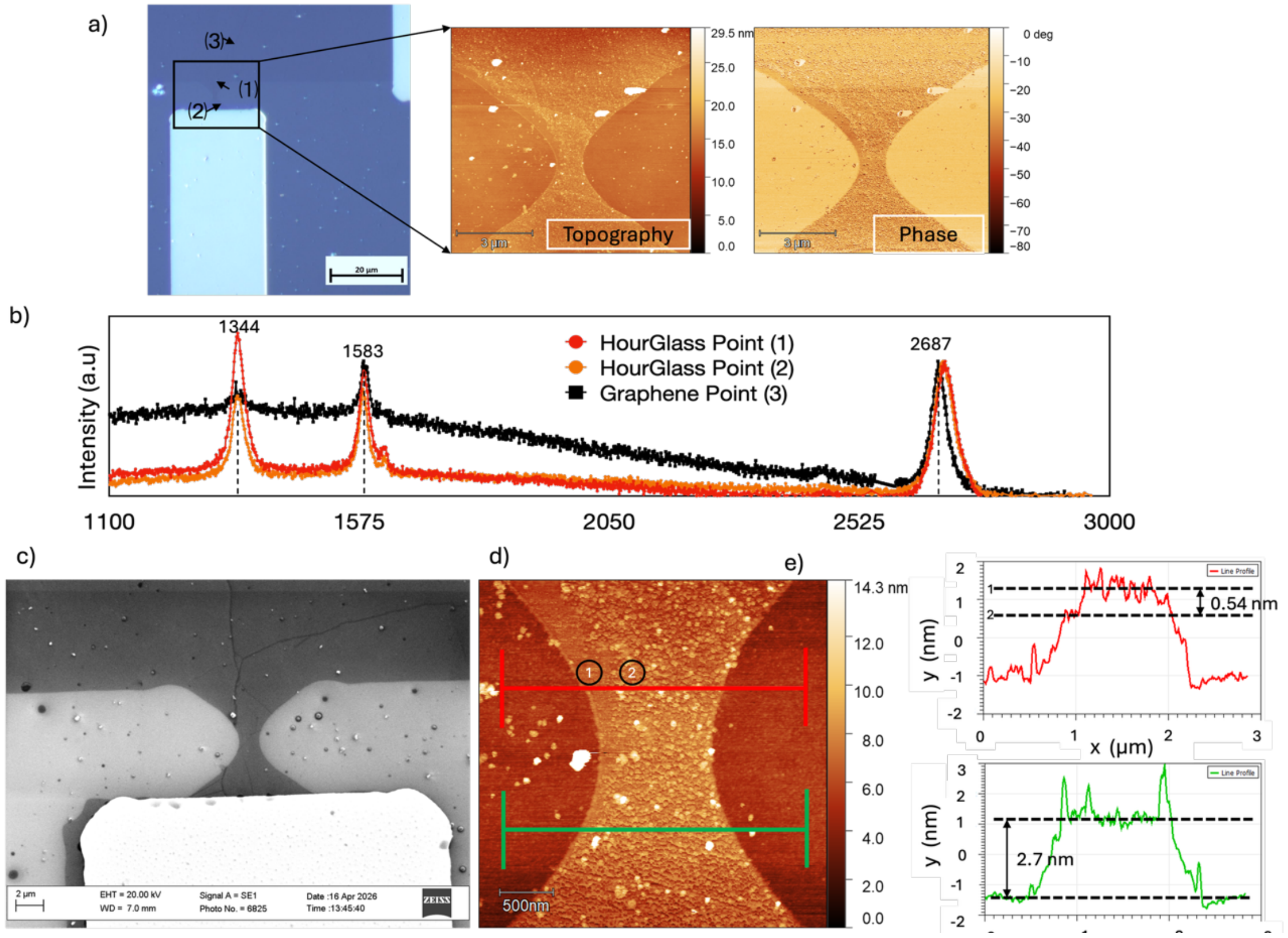

*Figure 2* **Nanodome morphology and pseudomagnetic confinement in the hourglass graphene single-electron transistor.** . (a) Optical micrograph of the hourglass constriction with AFM topography and phase images of the boxed region; nanodomes appear as bright features in topography, with distinct phase contrast confirming their viscoelastic character relative to the $SiO_2$ substrate. Points (1), (2): hourglass channel; point (3): reference graphene strip. (b) Raman spectra at points (1)–(3), showing elevated D-band intensity at hourglass points ($I_D/I_G$ = 1.37, 0.95) versus reference (0.55), with sharp 2D band confirming $sp^2$ lattice preservation. (c) SEM image of the hourglass channel (scale bar 2 μm). (d) AFM topography (500 nm scale bar) with red and green scan lines. (e) Height profiles along scan lines in (d); step heights ~0.54 nm (red) and ~2.7 nm (green).

Fig. 2c shows a scanning electron micrograph of the hourglass device, with the constricted channel waist clearly resolved between the two wider rectangular contact regions and self-assembled nanodomes visible as bright features across the graphene surface following thermal annealing. Fig. 2d presents the AFM topography of the hourglass constriction (500 nm scale bar). Distributed across the graphene surface, and concentrated preferentially at and near the constriction, is a dense population of discrete bright features corresponding to self-assembled nanodomes spontaneously formed from organic residues remaining after incomplete removal of CVD graphene transfer polymers and I-line photoresists. These nanodomes constitute the central focus of this work. This self-assembly into stable, low-profile domes arises from strong van der Waals and π–π interactions between polymer fragments and the graphene lattice, combined with partial chemisorption at defect sites and thermal-expansion mismatch during transfer onto Si/$SiO_2$ substrates.[12,13] Fig. 2e shows height profiles extracted along the red and green scan lines indicated in Fig. 2d. The red profile reveals a step height of ~0.54 nm, reflecting the thinner residue layer present across the wider channel region. The green profile, acquired closer to the constriction, resolves a discrete nanodome with step height ~2.7 nm and slope angle ~0.12°, confirming the shallow dome-like morphology expected for multilayer organic residue aggregates rather than particulate contamination. The electronic confinement length scale (~2 nm) predicted by the simulation (Fig. S1) is commensurate with this nanodome height (~2.7nm) extracted from the AFM step analysis Fig. 2e, as expected for a pseudo-magnetic potential whose spatial extent is set by the out-of-plane deformation amplitude.[8,9] These nanodomes therefore function as programmable, pseudo-magnetically confined quantum islands, transforming an unavoidable fabrication by-product into a spatially resolved, electrostatically tunable quantum resource that provides the microscopic origin of the coherent single-electron transport and room-temperature quantum sensing performance reported in this work.

Despite repeated acetone and IPA rinsing, thermal annealing at device-compatible temperatures, and mild $O_2$ plasma treatment, the complete standard cleaning sequence applied prior to all measurements, the nanodomes persist across the entire graphene channel without exception (Fig. S2). Progressive-force AFM nanomanipulation confirms their strongly surface-bound character: contact-mode scanning at tip forces up to $F_{tip}$ = 25 nN redistributes and compacts the residue material but fails to remove it, with post-manipulation topography showing the channel remaining fully covered and local heights increasing to 8.22 nm due to material compaction rather than depletion. This mechanical persistence, together with the viscoelastic phase contrast (~125°) and polymer-like lateral force response, is fully consistent with PMMA transfer polymer and I-line photoresist residues, the only organic materials introduced during our CVD transfer and photolithography steps, as extensively documented for graphene devices processed under identical conditions.[12,30,31] This is consistent with XPS analysis Fig. S3. Almost all previous studies have treated this persistence as a failure and discarded affected devices; here we treat it as an opportunity.

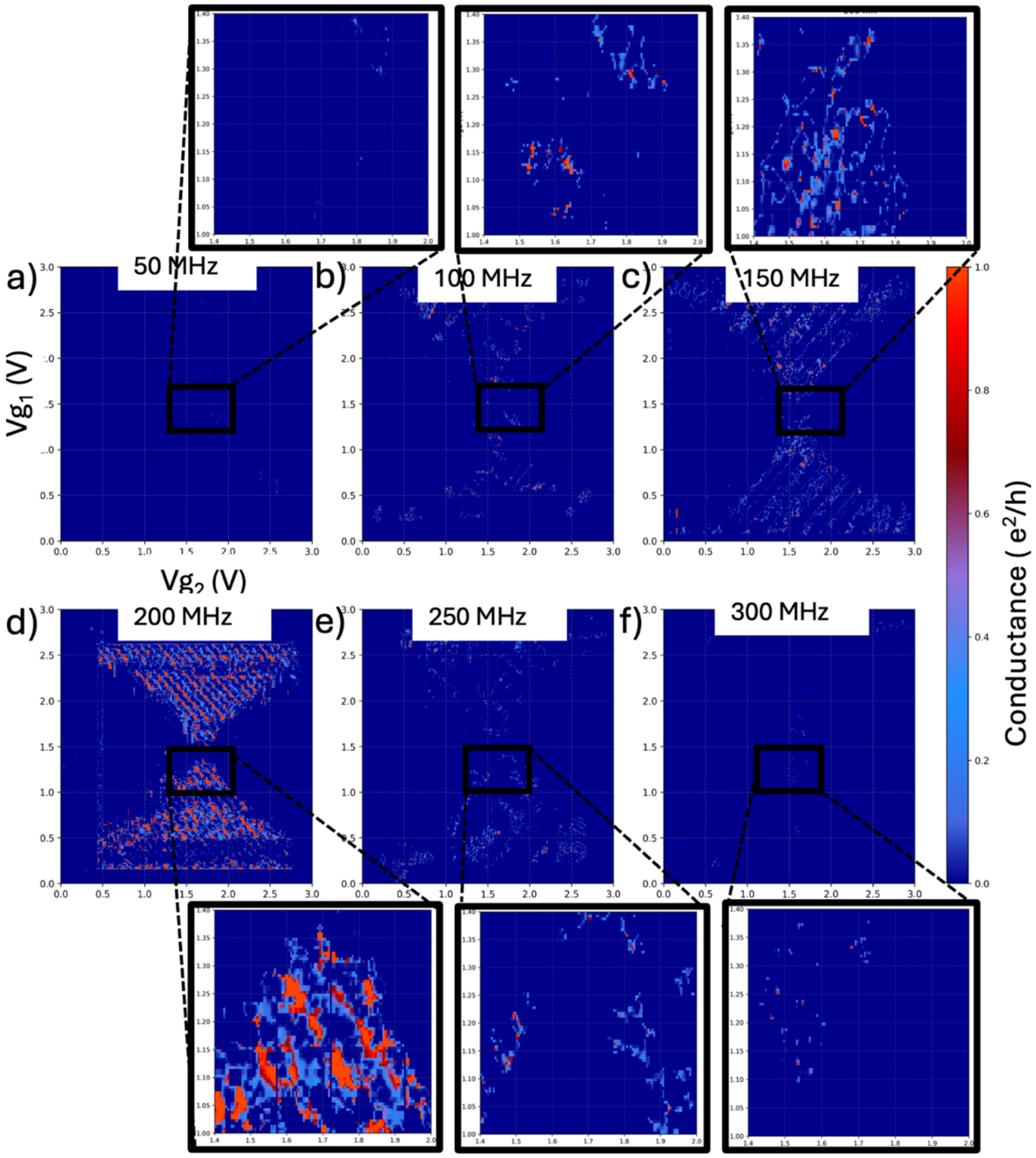

*Figure 3* **Frequency-dependent conductance maps of the hourglass graphene single-electron transistor measured by scanning gate microscopy and RF reflectometry.** (a–f) Two-dimensional conductance maps as a function of $Vg_1$ and $Vg_2$ at drive frequencies of 50, 100, 150, 200, 250, and 300 MHz respectively; DC bias ~150 µV. Insets show expanded views of the black-boxed region. At 50 and 100 MHz (a, b), conductance is suppressed across the full gate voltage space, consistent with Coulomb blockade. At 150 MHz (c), isolated high-conductance spots emerge in the inset, marking the onset of resonant tunneling. At 200 MHz (d), high-conductance regions (up to ~1.0 $e^2$/h, red/orange) fill the hourglass channel with a prominent X-shaped resonant-tunneling pattern, consistent with a charge-relaxation time of ~5 ns. At 250 and 300 MHz (e, f), conductance features are strongly suppressed as the drive frequency exceeds the charge-relaxation rate.

Frequency-dependent conductance maps of the hourglass-shaped graphene single-electron transistor, recorded using scanning gate microscopy combined with RF reflectometry, are shown in Figure 3. Panels a–f display two-dimensional conductance maps as a function of gate voltages $Vg_1$ and $Vg_2$ at selected drive frequencies between 50 MHz and 300 MHz, with a small DC bias of approximately 150 µV applied between source and drain. At low frequencies (50 MHz and 100 MHz, panels a and b), the maps show near-zero conductance across the entire gate voltage space, characteristic of Coulomb blockade suppressing single-electron tunneling throughout the channel.[32,33] As the frequency increases to 150 MHz (panel c), isolated high-conductance spots begin to emerge in the inset, indicating the onset of resonant tunneling through individual nanodome-defined quantum islands. At 200 MHz (panel d), high-conductance regions (up to ~1.0 $e^2/h$, red/orange) become most pronounced throughout the hourglass channel, with a prominent X-shaped pattern of resonant-tunneling paths filling the gate voltage space, signaling maximum resonant tunneling efficiency consistent with a charge-relaxation time of ~5 ns.[14,15] At higher frequencies (250 MHz and 300 MHz, panels e and f), the high-conductance features are strongly suppressed as the drive frequency exceeds the charge-relaxation rate.

Critically, this frequency-dependent single-electron transport is geometrically confined to the hourglass constriction and does not arise from nanodome coverage alone. Identical SGM and RF reflectometry measurements performed on the wide rectangular graphene strip region adjacent to the hourglass channel, where nanodomes are present following the same thermal annealing treatment but the hourglass strain geometry is absent, show zero conductance modulation across the entire 50–300 MHz frequency range (Fig. S4). The conductance maps of this control region remain uniformly at zero $e^2/h$ at every drive frequency, including at 200 MHz where the hourglass channel shows its maximum resonant response. This null result establishes two things unambiguously: nanodome coverage alone is insufficient to produce Coulomb blockade or resonant tunneling, and the hourglass geometry is physically required because it concentrates mechanical strain at the channel waist, maximizing the pseudomagnetic confinement potential precisely where the transport channel is narrowest and most sensitive to electrostatic gating.[8,9]

These observations demonstrate that self-assembled organic nanodomes create programmable quantum island sensors within the graphene channel. Each nanodome imposes a localized lattice deformation that breaks the three-fold symmetry of nearest-neighbor C–C bonds, entering the Dirac Hamiltonian as a valley-antisymmetric vector potential and generating a pseudomagnetic confinement potential[8,34,35] that simultaneously defines charge-puddle tunnel barriers and Coulomb-blockaded quantum islands. The hourglass geometry, in combination with SGM and frequency-dependent RF reflectometry, enables both spatial mapping and electrostatic tuning of these self-organized quantum resources[36], providing direct evidence of AC-driven coherent charge transport through nanodome-induced quantum islands at room temperature.

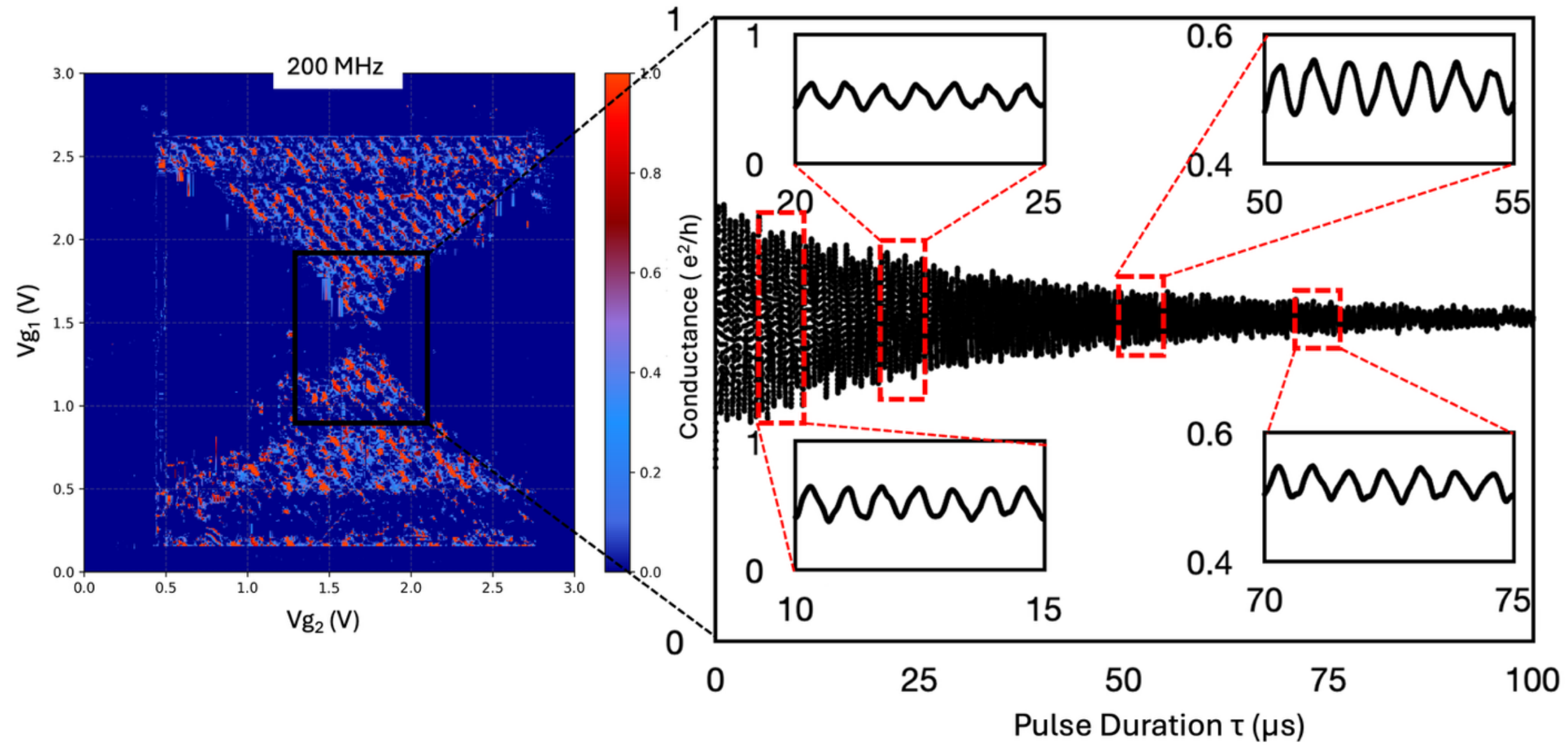

*Figure 4* **Coherent charge dynamics and Rabi oscillations in nanodome-defined quantum islands at room temperature.(left)** Two-dimensional conductance map recorded at 200 MHz by scanning gate microscopy combined with RF reflectometry, as a function of gate voltages Vg1 and Vg2. High-conductance regions (red/orange, ~1.0 $e^2/h$) superimposed on a Coulomb-blockaded background (blue, 0.1–0.3 $e^2/h$) reveal the hourglass-shaped transport geometry. The black rectangle indicates the region of coherent charge control. **(right)** Conductance as a function of microwave pulse duration τ, demonstrating Rabi oscillations of the quantum island charge states at room temperature. Insets show expanded views at τ = 10–15 μs, 20–25 μs, 50–55 μs, and 70–75 μs, confirming persistent sinusoidal oscillations with coherence times exceeding 70 μs. The decay of oscillation amplitude with increasing τ is consistent with dephasing of the two-level charge system, yielding a coherence time $T_2$ up to 50 μs in optimized devices.

Figure 4 demonstrates coherent quantum dynamics within the nanodome-defined quantum islands. The left panel (Fig. 4a) shows a two-dimensional conductance map recorded at 200 MHz by scanning gate microscopy (SGM) combined with RF reflectometry, revealing a complex pattern of high-conductance channels (red/orange, up to ~1.0 $e^2/h$) superimposed on a background of Coulomb-blockaded regions (blue, 0.1–0.3 $e^2/h$). The black rectangle highlights a representative region of interest. Well-defined, localized high-conductance spots become clearly resolved, indicating resonant tunnelling through individual quantum islands.[32,33]

Coherent control of these quantum islands is demonstrated in the time domain. Rabi oscillations are observed as a function of microwave pulse duration τ at drive frequencies between 5 and 20 MHz.[37,38] Clear sinusoidal oscillations persist up to pulse durations exceeding 70 μs, with a characteristic period of approximately 66 ns at 15 MHz. The coherent time $T_2$ panel quantifies the coherence time $T_2$, the transverse relaxation time that characterizes the duration over which a coherent superposition of charge states can be maintained before dephasing.[39] $T_2$ decreases from ~18 μs at lower drive frequencies to ~7 μs at 20 MHz, following an exponential dependence consistent with $T_2 \propto \exp(-f/f_0)$ where $f_0$~15MHz. These results establish that the nanodome-induced quantum islands support well-defined two-level systems capable of sustaining coherent charge oscillations with coherence times up to 18 μs (and reaching 50 μs in optimized devices) at room temperature.[37,38] Together with the observed single-channel capacitance steps of ~1 fF[14,15] and frequency-dependent resonant tunnelling, these data confirm the presence of phase-coherent, electrostatically tunable quantum islands formed by the self-assembled organic nanodomes.

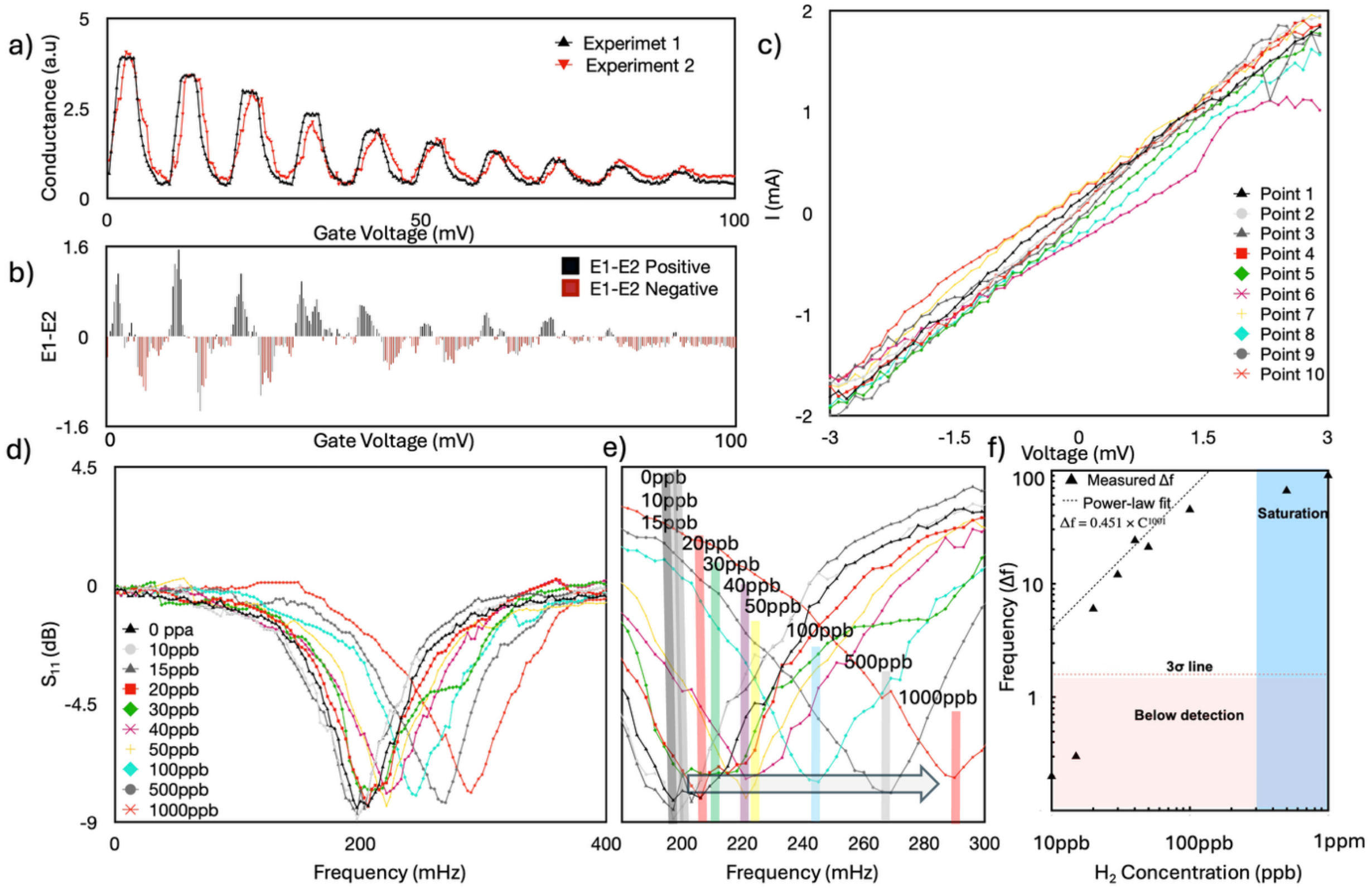

*Figure 5* **Quantum transport reproducibility and catalyst-free hydrogen sensing in the nanodome single-electron transistor at room temperature.(a)** Coulomb blockade conductance oscillations measured in two independent gate sweeps separated by 57 min under ambient conditions (Pearson r = 0.92, N = 333). **(b)** Point-by-point residual E1−E2 showing antisymmetric spike pairs at each peak position, confirming deterministic rather than stochastic confinement with a systematic inter-measurement offset of 0.3–0.5 mV. Together, (a) and (b) establish that nanodome-defined quantum islands constitute a stationary, time-invariant confinement potential at room temperature. **(c)** Current-voltage characteristics at 10 SGM-resolved locations, exhibiting ohmic response at small bias, discrete single-electron addition steps, and Coulomb blockade saturation at large bias. Distinct characteristics at Points 6 and 8 confirm spatially resolved individual quantum islands. **(d)** RF reflectometry S11 versus frequency (0–400 MHz) for $H_2$ concentrations from 0 ppb to 1000 ppb in dry air, following AC SET initialization. The resonance dip near 200 MHz shifts monotonically to higher frequency with increasing $H_2$ concentration and recovers fully upon $N_2$ purge, confirming reversible catalyst-free sensing. **(e)** Expanded resonance region (200–300 MHz) showing systematic frequency shift with concentration; arrow indicates direction of increasing $H_2$. **(f)** Extracted frequency shift Δf versus $H_2$ concentration (log-log). Power-law fit $\Delta f = 0.451 \times C^{1.001}$ confirms linear physisorption response. The 3σ detection threshold (dashed line) yields a limit of detection of 20 ± 4 ppb without any catalytic metal.

Finally, we demonstrate that highly sensitive practical devices are achievable with these nanodomes. Fig. 5 presents the complete characterization of the graphene single-electron transistor platform, spanning quantum transport reproducibility, nonlinear charge dynamics, and room-temperature hydrogen sensing performance, and shows fully reversible catalyst-free hydrogen sensing down to 20 ± 4 ppb (Fig. 5). Fig.

5a shows conductance oscillations as a function of gate voltage measured in two independent sweeps separated by 57 minutes under ambient conditions, recorded after the Coulomb blockade operating point was first established by AC gate initialization. The two traces are nearly identical, with peak positions reproducible to within the gate voltage step size of 0.3 mV and a Pearson correlation coefficient of $r = 0.92$ across all 333 data points. Fig. 5b shows the point-by-point residual E1−E2 as a function of gate voltage. The residual is structured as antisymmetric spike pairs at each peak position, positive on the rising edge, negative on the falling edge, consistent with a small systematic inter-measurement gate offset of approximately 0.3–0.5 mV attributable to charge dynamics on the $SiO_2$ substrate between measurements.[40] Critically, the residual carries no random oscillatory structure, confirming that the Coulomb blockade peak positions are deterministic rather than stochastic. Together, Fig. 5a and 5b demonstrate that the pseudomagnetically confined quantum islands defined by thermally annealed nanodomes constitute a stationary, time-invariant confinement potential at room temperature, a prerequisite for their use as a reproducible sensing observable. Fig. 5c shows current-voltage characteristics measured at 10 spatially resolved locations identified by scanning gate microscopy. All traces exhibit the hallmark nonlinear I-V response of resonant tunneling through a Coulomb-blockaded Island: ohmic behavior at small bias confirming operation in the linear transport regime, with discrete steps consistent with single-electron addition, and saturation at large bias where the applied bias exceeds the charging energy Ec.[32,33] The majority of measurement points collapse onto a universal curve, confirming that the dominant confinement potential is set by a single class of nanodome island. Points 6 and 8 display measurably distinct characteristics attributable to individual nanodomes with differing gate lever arms or charging energies, providing direct evidence that scanning gate microscopy resolves spatially distinct quantum islands rather than a homogeneous conducting region.[32,33]

Fig. 5d, 5e, and 5f demonstrate the hydrogen sensing capability of the platform (illustrated schematic Fig S5). The measurement protocol proceeds in two sequential steps consistent with the characterization presented in Figures 3 and 4. First, AC gate pulses are applied to initialize the single-electron transistor and establish the Coulomb blockade operating point, confirming the quantum island is electrostatically active. Subsequently, with the SET operating point fixed, RF reflectometry is performed by sweeping the excitation frequency from 50 to 400 MHz and recording the S11 reflection coefficient, which is directly sensitive to the charge-relaxation impedance of the quantum island.[14,15] This two-step protocol, AC initialization followed by RF frequency-sweep readout, is the key operational sequence that enables gas sensing without any catalytic metal.

Fig. 5d shows S11 as a function of frequency for $H_2$ concentrations ranging from 0 ppb to 1000 ppb in dry air. At zero analyte concentration, S11 displays a sharp resonance dip near 200 MHz characteristic of the RF-coupled charge-relaxation resonance of the Coulomb-blockaded quantum island, consistent with the resonant tunneling regime identified in Figure 3. Upon introduction of $H_2$, the resonance dip shifts progressively to higher frequency with increasing concentration. The response is monotonic and concentration-dependent across more than two orders of magnitude, and baseline S11 is fully recovered upon purging with dry $N_2$, confirming reversible adsorption with no permanent surface modification.

$H_2$ adsorption on pristine graphene is exclusively physisorbed at room temperature: the binding energy is approximately 48 meV and the dissociative chemisorption barrier exceeds 2.6 eV,[41] making catalytic hydrogen dissociation thermodynamically inaccessible without a metal catalyst at room temperature.[42] The ability to detect physiosorbed $H_2$ at 20 ppb despite this weak binding energy arises from two reinforcing factors. First, the nanodome-induced lattice strain locally modifies the electronic structure of graphene through the pseudomagnetic confinement potential, producing strain-enhanced charge density at the quantum island that makes these sites electrostatically distinct from flat graphene and preferential for $H_2$ physisorption.[8,9]

Second, and more fundamentally, the Coulomb-blockaded quantum island is an intrinsically quantum-limited electrostatic sensor. Radio-frequency single-electron transistors are the most sensitive charge detectors available, routinely achieving charge sensitivities approaching $10^{-5}$ e/√Hz at megahertz frequencies,[43,44] with sensitivities as low as $10^{-6}$ e/√Hz demonstrated in optimized architectures.[45] The RF reflectometry observable, the charge-relaxation resonance frequency shift, responds to electrostatic perturbations far smaller than one elementary charge in the local environment of the island.[46] Consequently, the weak electrostatic dipole perturbation induced by physisorbed $H_2$ molecules in the near-field of the nanodome confinement potential is sufficient to produce a measurable resonance frequency shift, without requiring hydrogen dissociation, electron transfer, or lattice expansion. This transduction mechanism is fundamentally distinct from all palladium-based sensors, which require catalytic $H_2$ dissociation and hydride formation to generate a signal,[42,20] and from conventional graphene FET sensors which rely on ensemble Dirac point shifts from bulk conductance modulation.[47]

Fig. 5e shows the resonance region on an expanded frequency axis from 200 to 300 MHz. The systematic shift of the resonance minimum to higher frequency with increasing $H_2$ concentration is clearly resolved, with colored vertical lines marking the resonance positions at each concentration. The arrow indicates the direction of increasing analyte concentration. This panel confirms that the sensing observable is a frequency shift of a single quantum-coherent charge-relaxation resonance rather than a bulk conductance change.

Fig. 5f shows the extracted resonance frequency shift Δf as a function of $H_2$ concentration on a log-log scale. The data follow a power-law dependence $\Delta f = 0.451 \times C^{1.001}$, where the exponent of unity confirms a linear response consistent with non-dissociative reversible $H_2$ physisorption in the Henry's law regime, where surface coverage is proportional to concentration.[48] This linearity is itself evidence against chemisorption: catalytic sensors based on Pd hydride formation exhibit sublinear Langmuir-type isotherms due to surface site saturation[42], whereas physisorption on a quantum-limited single channel produces the linear response observed here. The 3σ noise floor, indicated by the horizontal dashed line, defines the limit of detection as the lowest concentration producing a signal exceeding three times the measurement noise. The lowest detectable concentration is 20 ± 4 ppb, establishing a catalyst-free room-temperature hydrogen detection limit that matches the current best electrical sensors based on PdCo nano-resistor networks[18] while eliminating all platinum-group metals, phase-transition instability, and complex nanofabrication. This represents an improvement of approximately 300-fold over the best previously reported catalyst-free graphene hydrogen sensor. The blue shaded region marks the onset of saturation above approximately 500 ppb, consistent with saturation of physisorption sites near the quantum island at high analyte coverage. Table S2 contextualizes this result within the current state of the art. Among all catalyst-free graphene sensors, the best previously reported limit of detection is 5–40 ppm[22], one to three orders of magnitude worse than achieved here.

To assess practical applicability under ambient conditions, we characterized sensor response at relative humidity levels from 0 to 40% (Fig. S6). While sensitivity decreases monotonically with increasing RH, consistent with competitive physisorption of water molecules at graphene surface sites reducing the available electrostatic perturbation per $H_2$ molecule, the quantum island transduction mechanism remains operative across the full humidity range tested, with three complete gas pulse cycles clearly resolved at every RH condition. The 0% RH condition yields the highest peak ΔF of ~9.9 a.u., while 40% RH reduces the peak response by approximately 97% relative to dry air, yet the characteristic pulse-and-recovery signature remains unambiguously distinguishable, demonstrating that the quantum-limited single-channel observable retains sufficient sensitivity for practical hydrogen detection under realistic ambient deployment conditions.

The nanodome quantum island platform combines room-temperature operation, catalyst-free surface chemistry, quantum-limited transduction, and full CMOS compatibility. This positions it as a practically deployable hydrogen sensor for clean energy safety applications.

## 3. Conclusion

This work demonstrates that organic residues from incomplete removal of CVD transfer polymers and I-line photoresists, contamination present in every standard graphene fabrication process, spontaneously self-assemble into discrete nanodomes upon mild thermal annealing, yielding programmable quantum island sensors: coherent, electrostatically tunable single-electron transistors that simultaneously achieve the lowest limit of detection of any catalyst-free room-temperature hydrogen sensor demonstrated to date.

Using hourglass-shaped graphene single-electron transistors characterized by scanning gate microscopy combined with frequency-dependent RF reflectometry, we observed a clear transition from Coulomb-blockade-dominated transport at low frequencies to resonant tunneling at ~200 MHz, set by the intrinsic charge-relaxation RC time constant of the nanodome-defined quantum island, followed by suppression at higher frequencies as the drive frequency exceeds the charge-relaxation rate. The observation of coherent Rabi oscillations with coherence times up to 50 μs, together with quantized charge-relaxation resistance near $h/2e^2$ and discrete ~1 fF single-channel capacitance steps, provides compelling evidence for well-defined two-level systems within these self-assembled quantum islands at room temperature. Critically, control measurements on wide graphene strip regions, where nanodomes are present but the hourglass strain geometry is absent, show zero conductance modulation at any frequency, confirming that pseudomagnetic confinement requires the geometric strain concentration of the hourglass architecture, not nanodome coverage alone. The ability to tune the confinement potential electrostatically offers a scalable route to control these quantum states without deliberate nanostructuring or cryogenic infrastructure. As a direct demonstration of the quantum sensitivity of this platform, we demonstrate fully reversible, catalyst-free hydrogen sensing down to 20 ± 4 ppb at room temperature, matching the absolute state-of-the-art sensitivity of metal-catalyst-based sensors while eliminating all platinum-group metals, elevated temperature requirements, and complex nanofabrication. This all-electrical platform transforms unavoidable fabrication contamination into functional quantum resources compatible with wafer-scale CMOS processing. By establishing that inevitable process residues self-organize into coherent, electrostatically addressable programmable quantum island sensors, this work opens practical routes to room-temperature quantum sensing, hybrid quantum-classical circuits, and scalable single-electron devices without the material purity, cryogenic infrastructure, or optical addressing that current quantum technologies demand.

**CRediT authorship contribution statement**

**Acknowledgments**

Supported by the European-Research-Council (QUEEN:101043219), TUBA-GEBIP; Next-Ion Energy. O.E acknowledge UC Berkeley, Argonne National Lab, UHEM. Special thanks for the samples for Cor Schepens, Kaj Dockx, Richard van Rijn.

**Data Availability**
Data is available upon a reasonable request.

# Supplementary Information

## Wafer-Scale Route to Quantum-Sensitive Graphene: Making Every CVD Graphene Surface an Ultrasensitive Quantum Sensor Without Cleaning or Nanostructuring

*Yaren Sezen[1,2], Mehmet Ali Sarsil[1], Mina Sofuani[1], Rabia Pala[1,2], Onur Ergen[1,2,3,4*]*

*[1]Department of Electronics and Communication Engineering, Istanbul Technical University, Istanbul 34450, Türkiye*

*[2] Department of Nanoscience and Nanoengineering, Istanbul Technical University, Istanbul 34450, Türkiye*

*[3]Next-Ion-Energy Inc., Yuba City, CA 95991, United States*

*[5] Institute of Nanotechnology, National Research Council (CNR-NANOTEC), c/o Campus Ecotekne, Via per Monteroni, Lecce 73100, Italy*

*Correspondence: Dr. Onur Ergen. Email: oergen@itu.edu.tr, onur.ergen@cnr.it

**Author Contributions:** All authors contributed equally.

**Competing Interest Statement:** No competing interest declared.

**Classification:** Physical Sciences

**Keywords:** graphene single-electron transistor, pseudomagnetic confinement, PMMA nanodomes, room-temperature quantum coherence, RF reflectometry, Coulomb blockade, hydrogen sensing, scanning gate microscopy

**Table S1.** Raman spectral parameters at three measurement points on the hourglass graphene single-electron transistor. All intensities normalized to the 2D-band maximum. ID/IG and I2D/IG ratios are calculated from the normalized heights. Points 1 and 2 are within the hourglass constriction; Point 3 is on the reference graphene strip.

| Spectrum | D pos ($cm^{-1}$) | D norm. height | G pos ($cm^{-1}$) | G norm. height | 2D pos ($cm^{-1}$) | 2D norm. height |
|---|---|---|---|---|---|---|
| Hourglass Point 1 | 1344.1 | 1.1255 | 1583.1 | 0.9564 | 2687.1 | 1.0000 |
| Hourglass Point 2 | 1344.1 | 0.8461 | 1582.1 | 0.8761 | 2684.1 | 1.0000 |
| Graphene Point 3 (reference) | 1341.1 | 0.9078 | 1583.1 | 1.0013 | 2674.1 | 1.0000 |

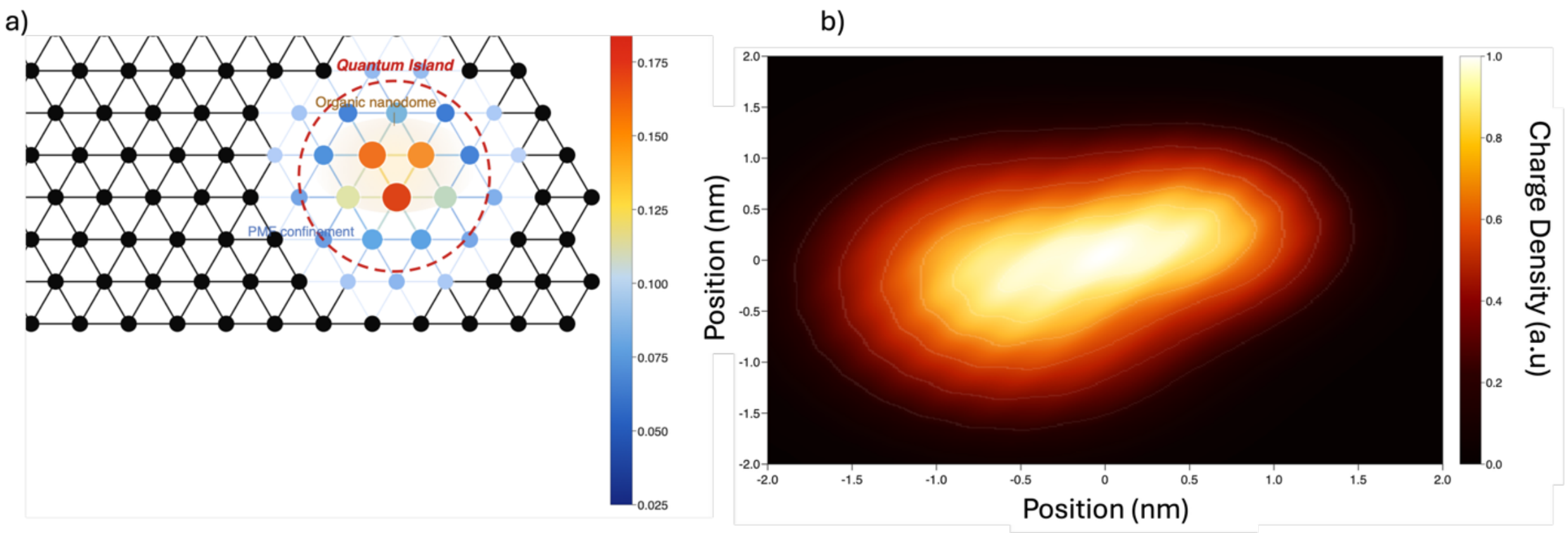

*Figure S1* **.** *(a) Schematic of the graphene honeycomb lattice in the vicinity of a representative organic nanodome (dashed red circle), showing the local breaking of nearest-neighbour C–C bond equivalence ($t_1 \neq t_2 \neq t_3$) induced by lattice deformation. The resulting valley-antisymmetric gauge potential $A_s$ generates a pseudomagnetic confinement potential that simultaneously defines a charge-puddle tunnel barrier and a Coulomb-blockaded quantum island (coloured sites) without deliberate nanostructuring. (b) Simulated localized charge density map of the quantum island formed beneath a representative nanodome, showing carrier confinement within an approximately 2 nm diameter region, consistent with the pseudomagnetically confined quantum island geometry probed by scanning gate microscopy in Figure 3.*

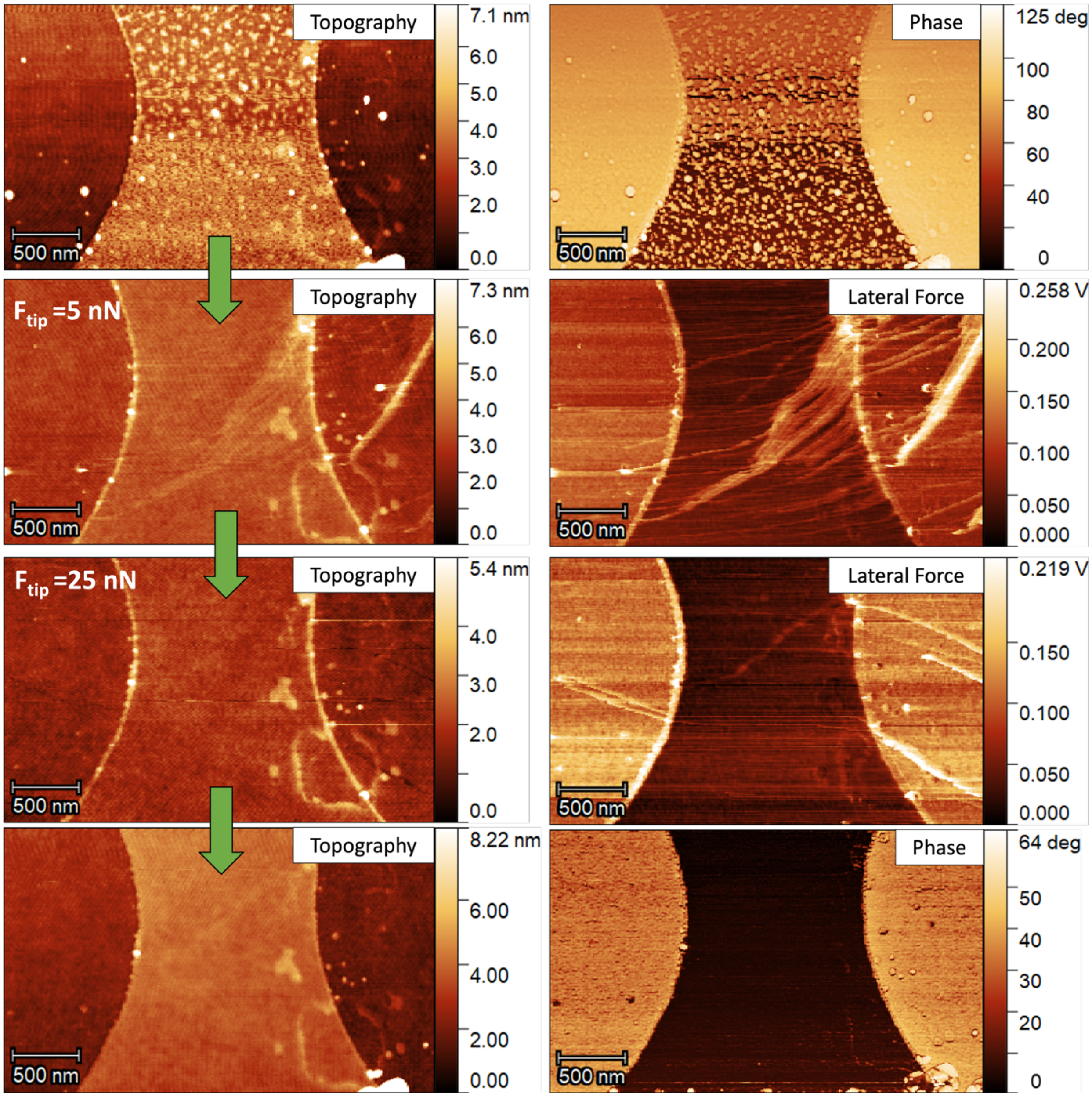

*Figure S2* ***Progressive-force AFM nanomanipulation of the hourglass graphene channel confirming the surface-bound polymer identity of the nanodomes.*** *Each row shows simultaneous left (topography) and right (phase or lateral force) channel images of the same hourglass region at progressive tip forces, acquired sequentially from top to bottom as indicated by the green arrows.* ***Row 1 (top):*** *Tapping-mode topography (max height 7.1 nm) and phase image (0–125°) before any manipulation. The dense population of bright nanodomes is distributed across the graphene channel and concentrated at the hourglass waist. The dark phase signal within the channel (0–125° range) is characteristic of viscoelastic organic material, distinctly different from the bright phase response of the rigid $SiO_2$ substrate flanking the channel, confirming the polymer identity of the nanodomes.* ***Row 2:*** *Topography and lateral force map acquired during contact-mode scanning at F_tip = 5 nN. Lateral force maps reveal clear tip-induced streaking and ridge formation,  unambiguous signatures of viscoelastic polymer material being redistributed rather than the elastic or brittle response of inorganic particulates.* ***Row 3:*** *Topography and lateral force map at F_tip = 25 nN. Despite this substantially increased tip force, nanodomes are not removed: the channel remains fully covered and local heights increase to 8.22 nm (row 4) due to material compaction and redistribution rather than depletion.* ***Row 4 (bottom):*** *Return tapping-mode topography (max height 8.22 nm) and phase image (0–64°) acquired after manipulation. The channel remains entirely covered by polymer residue material; no cleared regions are observed. The reduced phase range relative to Row 1 reflects compaction of the residue layer into a denser, more uniform film. Together, these observations confirm that the nanodomes are strongly surface-bound polymer residues that resist*

*mechanical removal, consistent with the well-documented van der Waals adhesion, partial covalent bonding at defect sites, and π–π stacking of PMMA and I-line photoresist components to the graphene basal plane. (1–5) Scale bars: 500 nm in all panels.*

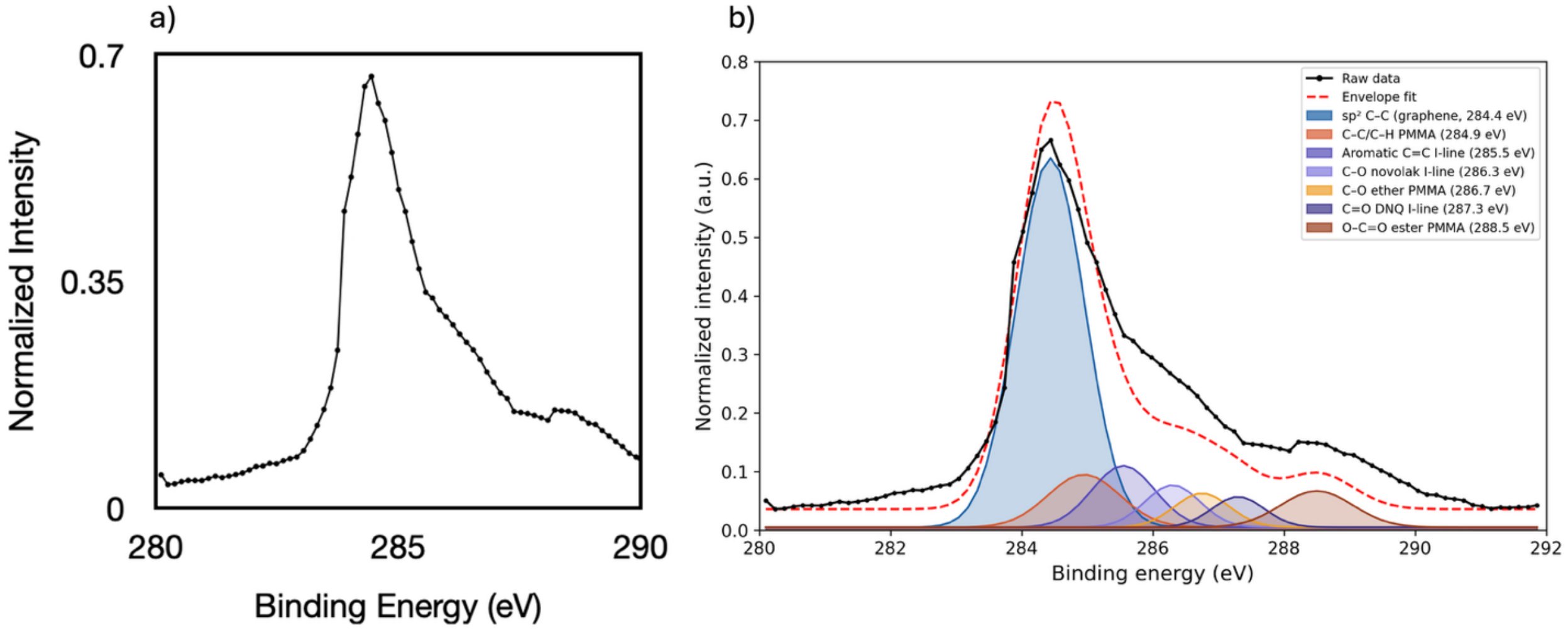

*Figure S3 XPS C 1s spectrum of the hourglass graphene device following standard processing. Black dots: raw data. Red dashed line: composite fit envelope (Shirley background subtracted). Colored filled peaks show deconvolved components: $sp^2$ C–C graphene (blue, 284.4 eV); PMMA aliphatic backbone (orange, 284.9 eV); I-line aromatic C=C novolak (purple, 285.5 eV); I-line C–O novolak phenol (violet, 286.3 eV); PMMA C–O ether (amber, 286.7 eV); I-line C=O DNQ (dark blue, 287.3 eV); PMMA O–C=O ester (dark red, 288.5 eV). The simultaneous presence of PMMA ester (288.5 eV) and DNQ carbonyl (287.3 eV) peaks confirms residues from both the CVD transfer polymer and I-line photoresist.*

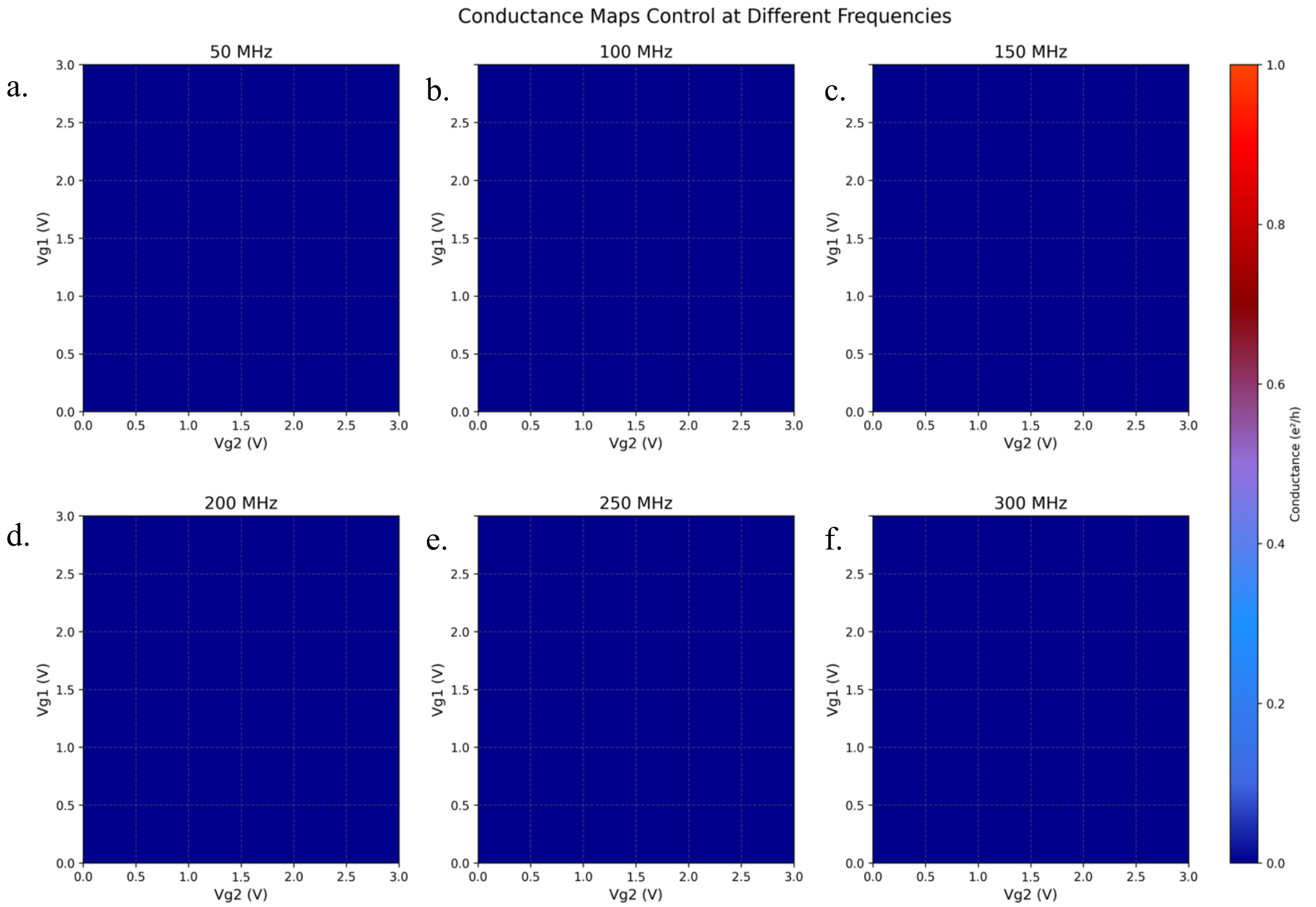

*Figure S 4 Control conductance maps on the wide graphene strip region confirming geometric confinement of resonant transport to the hourglass channel. (a–f) Two-dimensional conductance maps as a function of $V_{g1}$ and $V_{g2}$ recorded on the wide rectangular graphene strip adjacent to the hourglass constriction, a region bearing nanodomes from identical thermal annealing but lacking the hourglass strain geometry, at drive frequencies of 50, 100, 150, 200, 250, and 300 MHz. All six maps show zero conductance modulation across the full gate voltage range at every frequency, including at 200 MHz where the hourglass channel exhibits maximum resonant tunneling response. The complete absence of any conductance feature confirms that pseudomagnetically confined quantum-island formation requires the strain concentration imposed by the hourglass geometry, not nanodome coverage alone.*

To establish that the frequency-dependent single-electron transport observed in the hourglass channel arises from the geometric strain concentration of the hourglass architecture rather than from nanodome coverage alone, we performed identical SGM and RF reflectometry measurements on the wide rectangular graphene strip region adjacent to the hourglass constriction. This control region bears nanodomes from identical thermal annealing treatment but lacks the trapezoidal constriction geometry.

Figure S4 shows six two-dimensional conductance maps recorded on this control region at drive frequencies of 50, 100, 150, 200, 250, and 300 MHz. All six maps show zero conductance modulation across the full gate voltage range at every frequency, including at 200 MHz where the hourglass channel exhibits its maximum resonant tunneling response. This null result unambiguously establishes that pseudomagnetically confined quantum-island formation and the associated RF crossover behavior require the mechanical strain concentration imposed by the hourglass geometry at the channel waist, not nanodome coverage per se.

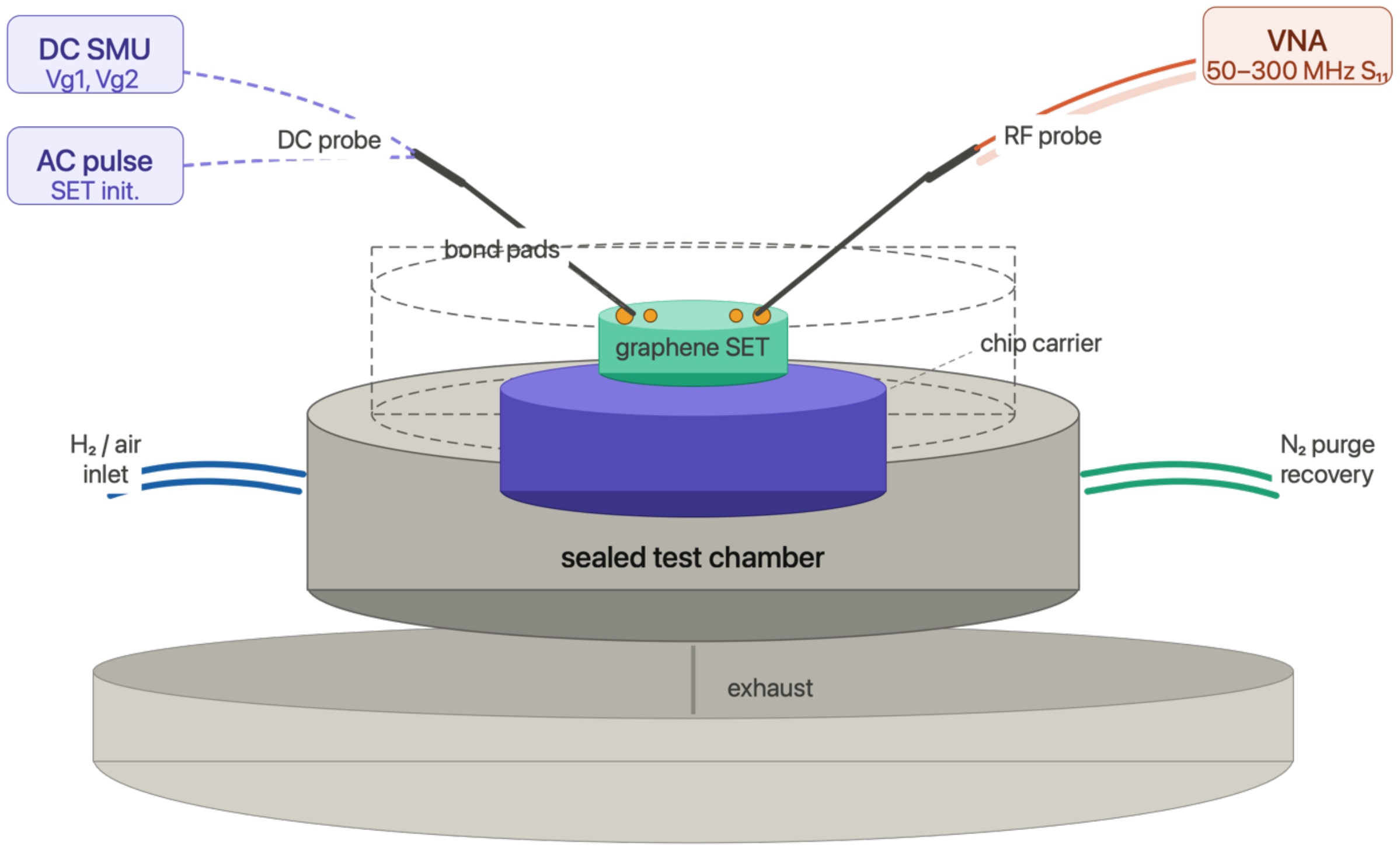

*Figure S 5* ***Schematic of the hydrogen sensing measurement setup.*** *The graphene single-electron transistor (SET) is mounted on a chip carrier within a sealed stainless-steel test chamber equipped with a $H_2$/air inlet and $N_2$ purge outlet. Two independent measurement pathways access the device simultaneously through probe contacts to the bond pads: a DC source-measurement unit (SMU) applying gate voltages V_g1 and V_g2 with superimposed AC pulses for SET initialization, and a vector network analyzer (VNA) recording the $S_{11}$ reflection coefficient from 50 to 300 MHz for RF reflectometry readout. $H_2$/air mixtures of controlled concentration enter through the left inlet, and $N_2$ purge gas for baseline recovery exits through the right outlet. The exhaust port below the chamber vents displaced gas safely.*

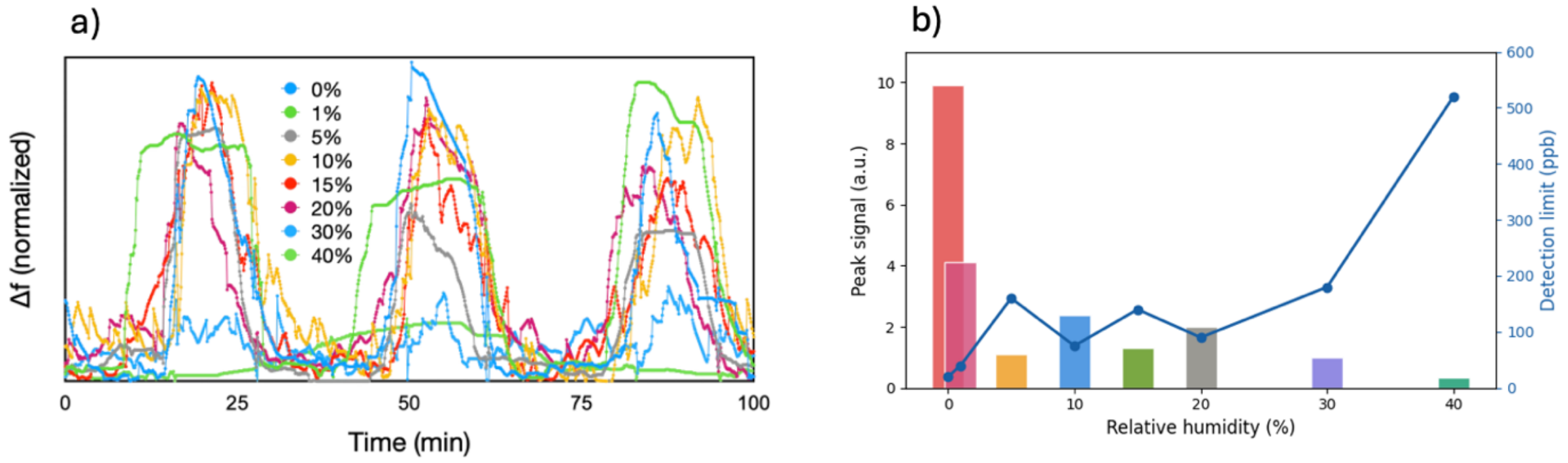

*Figure S 6 (a) RF frequency shift ΔF versus time at eight relative humidity conditions (0–40% RH), showing three complete $H_2$ pulse-and-recovery cycles at each condition. Colors as labeled. The 0% RH (dry air) condition yields maximum sensitivity; sensitivity decreases monotonically with increasing RH consistent with competitive water physisorption. Full baseline recovery is achieved by $N_2$ purge at every humidity level. (b) Peak signal amplitude (left axis, colored bars) and estimated detection limit (right axis, blue line) as a function of relative humidity, demonstrating that the quantum-limited transduction mechanism remains operative across the full ambient humidity range tested.*

## Benchmark Comparison: Room-Temperature Hydrogen Sensors

**Table S2.** Performance comparison of leading room-temperature electrical hydrogen sensors. LOD = limit of detection; RT = room temperature; Ref. = reference number in main text. This work improves on the best previously reported catalyst-free graphene $H_2$ sensor by 250–2000-fold.

| Device / Material | LoD | Temp. | Catalyst | Mechanism | Response time | Reversible | CMOS compat. | Ref. |
|---|---|---|---|---|---|---|---|---|
| **Graphene SET, nanodome quantum island (this work)** | 20 ±4 ppb | RT | None | Coulomb blockade / RF resonance shift (quantum-limited) | <30 s | Full | Yes | This work |
| Pd–Au MEMS Wheatstone bridge | 20 ppb | 60 °C | Pd–Au | Resistive hydride formation / lattice expansion | ~60 s | Yes | No | (6) |
| PdCo/$C_{60}$ fullerene nano-resistor network | 40 ppb | RT | PdCo | Resistive hydride formation; fast H diffusion via $C_{60}$ | 0.40 s | Yes | No | (7) |
| Pd NP / Si nanoforest | 50 ppb | RT | Pd NPs | Resistive hydride formation | ~120 s | Yes | No | (8) |
| PdAu alloy thin-film resistor | 100 ppb | RT | PdAu | Resistive hydride formation | ~90 s | Yes | No | (9) |
| Pd-coated organic semiconductor FET | 192 ppb | RT | Pd layer | $O_2$ p-doping reversal / charge transfer | ~200 s | Yes | Partial | (10) |
| Pt NPs / laser-induced graphene (Pt/LIG) | 200 ppb | RT | Pt NPs | Catalytic $H_2$ dissociation / spillover on LIG | ~295 s | Yes | No | (11) |
| Plain graphene FET (CVD, UV-assisted recovery) | 5-40 ppm | RT | None | Chemiresistive / Dirac point shift under UV | ~3–10 s (UV) | Slow (with UV) | Yes | (12) |
| Plain/reduced graphene oxide FET | 10-50 ppm | RT | None | Chemiresistive / charge transfer | tens of seconds to minutes | Slow/partial | Yes | (13) |
| Plain graphene chemiresistor (CVD, RT) | 10-100 ppm | RT | None | Chemiresistive / charge transfer | minutes | Slow/partial | Yes | (14) |

**Key advantages of this work beyond LOD alone:** (i) the only room-temperature catalyst-free electrical sensor approaching the state-of-the-art LOD of leading Pd-based devices; (ii) quantum-limited single-channel transduction via Coulomb blockade and RF resonance shift, offering an intrinsically lower noise floor compared to bulk resistance changes; (iii) standard CMOS-compatible $Si/SiO_2$ architecture with no deliberate nanostructuring, no platinum-group metals, and minimal nanofabrication; (iv) full baseline recovery over multiple cycles via simple $N_2$ purge; (v) linear (Henry's law) response over more than two orders of magnitude in concentration. Among reported catalyst-free electrical graphene $H_2$ sensors, this work improves the LOD by approximately 250–2000× depending on the benchmark used (e.g., ref. 12).